


\magnification 1200
\hfuzz=7pt



\def\h{{\cal H}}

\def\o{O(3,2)}

\def\C{\hbox{\bf C}}
\def\R{\hbox{\bf R}}
\def\P{\hbox{\bf P}}
\def\Z{\hbox{\bf Z}}

\def\ovk{\overrightarrow{\kern-2pt K\kern2pt}}
\def\ovp{\overrightarrow{\kern-2pt P\kern2pt}}
\def\ovs{\overrightarrow{\kern-2pt S\kern2pt}}
\def\ovj{\overrightarrow{\kern-2pt J\kern2pt}}
\def\ovx{\overrightarrow{\kern-2pt X\kern2pt}}

\def\demi{{1\over 2}}

\def\kz{$\k\rightarrow0$}

\font\fiveti=cmmi5
\def\Osp{OSp(1/2)}
\def\osp{osp(1/2)}
\def\su{SU(1,1)}
\def\ssu{su(1,1)}

\def\N{\hbox{\bf N}}
\def\E{\hbox{\bf E}}
\def\F{\hbox{\bf F}}

\def\om{{\cal O}_M}
\def\am{{\cal A}_M}
\def\nil{{\cal N}}
\def\she{{\cal E}}
\def\d{{\cal D}^{(1)}}
\def\dd{{\cal D}^{(1|1)}}
\def\mz{|z|^2}
\def\zb{\bar z}
\def\ko{|0\rangle}
\def\bo{\langle 0}
\def\kz{|z\rangle}
\def\bz{\langle z}
\def\pz{{\partial\over\partial z}}
\def\pzb{{\partial\over\partial \zb}}
\def\t{\theta}
\def\tb{\bar \theta}
\def\kzt{|z,\theta\rangle}
\def\bzt{\langle z,\theta}
\def\pt{{\partial\over\partial \theta}}
\def\ptb{{\partial\over\partial \tb}}


\def\picture #1 by #2 (#3){
        \vbox to #2{
             \hrule width #1 height 0pt depth 0pt
              \vfill
                                                        \special{picture #3}}}

\def\scaledpicture #1 by #2 (#3 scaled #4){{
        \dimen0=#1  \dimen1=#2
        \divide\dimen0 by 1000  \multiply\dimen0 by #4
        \divide\dimen1 by 1000  \multiply\dimen1 by #4
        \picture  \dimen0 by \dimen1 (#3 scaled #4)}}

\def\ref#1{\noindent\llap{[{\bf #1}]\quad}}
\def\lmp#1{\noindent\llap{{\bf #1}\quad}}
\def\refe#1{\item{\hbox to\parindent{\enskip[{\bf #1}]\hfill}}}
\def\boxit#1#2{\vbox{\hsize 8cm\hrule\hbox{\vrule\vbox spread #1{\vfil\hbox
spread #1{\hfil#2\hfil}\vfil}%
\vrule}\hrule}}
\def\page#1{\leaders\hbox to 5mm{\hfil.\hfil}\hfill
\rlap{\hbox to 5mm{\hfill#1}}\par}

\font\sect=cmbx10 scaled\magstep2

\newif\ifpagetitre                          \pagetitrefalse
\newtoks\hautpagetitre                  \hautpagetitre={\hfil}
\newtoks\chapitrecourant               \chapitrecourant={\hfil}
\newtoks\titrecourant                     \titrecourant={\hfil}
\newtoks\hautpagegauche                \hautpagegauche={\hfil}
\newtoks\hautpagedroite
\hautpagedroite={\hfil\the\titrecourant\hfil}
\headline={\ifpagetitre\the\hautpagetitre
                  \else\ifodd\pageno\the\hautpagedroite
                  \else\the\hautpagedroite\fi\fi}
\footline={\hfil\bf\folio\hfil}

\def\makeheadline{\vbox to 0pt{\vskip -30pt
       \line{\vbox to8.5pt{}\the\headline}\vss}\nointerlineskip}
\def\makefootline{\baselineskip=35pt\line{\the\footline}}
\vsize=7.2 in
\voffset=0 mm
\hsize=138 mm
\baselineskip=16pt
\parskip=4pt plus 1pt minus 1pt

\line{January 29, 1993\hfill  CRM-1850}
\line{ \hfill }
\vglue 1cm
\centerline{\sect On the Supersymplectic Homogeneous
Superspace}
\medskip
\centerline{\sect Underlying the OSp(1/2) Coherent States}
\vglue 1cm
\centerline{\bf Amine M. El Gradechi\footnote{ *}{\it
e-mail: elgradec@ERE.UMontreal.CA}}
\vglue 0.1cm
\centerline{\it Centre de Recherches Math\'ematiques, Universit\'e de
Montr\'eal}   \centerline{\it C.P. 6128-A, Montr\'eal (Qu\'ebec) H3C 3J7,
Canada} \smallskip \centerline{\it and}
\smallskip
\centerline{\it Department of Mathematics, Concordia University}
\centerline{\it Montr\'eal (Qu\'ebec) H4B 1R6, Canada}
\vglue 1cm
\centerline{\bf ABSTRACT}
\vglue 0.2cm
In this work we extend Onofri and Perelomov's coherent states methods to
the recently introduced $\Osp$ coherent states.  These latter are shown to be
parametrized by points of a supersymplectic supermanifold, namely the
homogeneous superspace $\Osp/U(1)$, which is clearly identified with a
supercoadjoint orbit of $\Osp$ by exhibiting the corresponding equivariant
supermoment map.  Moreover, this supermanifold is shown to be a nontrivial
example of Rothstein's supersymplectic supermanifolds.  More precisely,
we show that its supersymplectic structure
is completely determined in terms of $SU(1,1)$-invariant (but unrelated)
K\"ahler $2$-form and K\"ahler metric on the unit disc.
This result allows us to define the notions of a superK\"ahler supermanifold
and a superK\"ahler superpotential, the geometric structure of the former
being encoded into the latter.

\vfill\eject
\noindent{\bf 1. Introduction}
\medskip
\noindent In the past thirty years several Lie algebraic, group theoretic and
geometric concepts have been successfully extended to the supersymmetric
context, enriching these already established mathematical structures
and giving birth to a new branch of mathematics called supermathematics.
Numerous review articles and textbooks are available, see for instance [1-5].
For applications in physics we refer to [6].

Recently Balantekin {\sl et al.} [7-8] (see also [9]) went a step further
and introduced the so called supercoherent states (SCS), which extend in a
natural way the well known Perelomov coherent states (CS) [10].  It seems
then natural to pursue further this parallel with the non-super case, by
addressing such questions as: {\it what is the geometric structure
underlying these SCS\/}?  In order to answer this question we will
devote the first part of this work to the extension of Onofri's analysis
[11].  Let us briefly recall the main features of this analysis.  Perelomov's
coherent states for a semi-simple Lie group $G$ are points of an orbit of a
unitary irreducible representation (UIR) $U$ of $G$ in a Hilbert space $\h$
(more
precisely in $\C\P(\h)$).  Choosing an initial state in $\h$, called the
fiducial
state, the vectors of the corresponding $G$-orbit in $\h$ are parametrized by
points of a homogeneous space $G/H$, where $H$ is the isotropy
subgroup (up to a phase) of the fiducial state.  Using this family of states
(CS)
Onofri showed that $G/H$ is a symplectic and even a K\"ahler manifold [11].
More precisely, he showed that a symplectic structure on $G/H$ can be
explicitly obtained from the coherent states.  From the quantization vs.
classical limit point view, Onofri's result identifies, through the CS, the
classical
limit of
the quantum theory described by the UIR $(U, \h)$ of $G$.

Here, we shall answer the above question in the case of the $\Osp$ CS
constructed in [7].  In fact, we will show that the homogeneous
superspace $\Osp/U(1)$ parametrizing these SCS is a supersymplectic
supermanifold [1] [4].   Indeed, using an extension of Onofri's analysis we
will exhibit a supersymplectic structure on that homogeneous superspace.
Moreover, using a supersymmetric extension of Berezin's covariant symbols
[12], we will be able to construct an equivariant supermoment map
allowing us to show that $\Osp/U(1)$ is a supercoadjoint orbit
of $\Osp$.

The general notion of a supersymplectic supermanifold has been considered
by both Berezin [1] and Kostant [4] using different conventions.  Since the
second reference gives a complete description of the subject, we will use
its conventions.  Recently, in extending Batchelor's theorem [13] which
allows the description of supermanifolds in terms of usual geometry,
Rothstein [14] proved a similar result for the
supersymplectic subcategory.  In fact, according to Rothstein any
supersymplectic supermanifold can be completely characterized by a set
$(M, \omega, \E, g, \nabla)$, where $(M, \omega)$ is a symplectic manifold
and $\E$ is a vector bundle over $M$ with a metric $g$ and a $g$-compatible
connection $\nabla$.  In the second part of this work, we will explicitly
identify
these five ingredients
for the supersymplectic supermanifold underlying the $\Osp$ CS,
exhibiting thus a nontrivial example of Rothstein's theorem.  Moreover, from
this
example we will be able to define the new notion of a superK\"ahler
supermanifold.

The notion of a K\"ahler potential appears naturally in Onofri's analysis.  It
constitutes the crucial step of that analysis, since it is defined
in terms of the CS and it gives rise to the K\"ahler structure on the
homogeneous space parametrizing these CS.  Hence, it connects the quantum
model to its classical limit. We shall show here that this notion
extends also to the super-setting, allowing one to encode Rothstein's
theorem ingredients for a superK\"ahler supermanifold into a superK\"ahler
superpotential.

Since $\Osp$ is the simplest supersymmetric extension of $SU(1,1)$ we
shall start this letter in section 2 by rederiving Onofri's analysis for the
$SU(1,1)$ CS.  In section 3 we will extend it to $\Osp$ basing our
construction on the SCS obtained in [7].  We will exhibit the
supersymplectic form and show that $\Osp/U(1)$ is a supercoadjoint orbit
of $\Osp$.  Then in section 4, we will show that what we have obtained is a
non-trivial example of Rothstein's supersymplectic supermanifolds.  We will
also define the notion of a superK\"ahler supermanifold and show that the
ingredients of Rothstein's theorem can be read off from a superK\"ahler
superpotential.  In section 5 we will show that our supersymplectic
supermanifold satisfies the Darboux-Kostant theorem [4].  Finally, section
6  gathers a discussion, concluding remarks and prospects.  Notice that
throughout this work, $\Osp$ stands for $OSp(1/2,\R)$.

\bigskip
\noindent{\bf 2. SU(1,1) CS and the symplectic unit disc} \medskip
\noindent
Coherent states are special quantum states that have proven to be very
useful in many areas of physics, especially in quantum optics (see [15] for
references).  They have also proven to be very useful in mathematical
physics, mainly in connection with the quantization vs. classical limit
procedures [11] [12] [15].  As stressed in the introduction this last aspect
will be our focus here.  The group theoretic construction of coherent states
[10] holds for Lie groups possessing square integrable representations.
Hence for non-compact semisimple Lie groups such CS live within the
discrete series representations.  Let us consider here the simplest example
of the $\su$ CS.

The Lie algebra $\ssu$ of $\su$ is described by the following commutation
relations in the Cartan-Weyl basis,
$$[K_0, K_\pm]=\pm K_\pm\quad\hbox{and}\quad[K_+,
K_-]=-2K_0.\eqno(2.1)$$
The positive discrete series representations of $\su$ are denoted $D^k_+$
for $k\in\demi\N$ and $k>\demi$.  For given $k$, the representation
space is spanned by the set of states $\{\, |k,m\rangle$, $m=k, k+1, k+2,
\ldots\}$.  These basis elements are eigeinstates of $K_0$ with
eigenvalues $k+m$.  They result from the action of integer powers of $K_+$
on the ``highest weight'' state $|k, k\rangle$ defined by
$$K_0|k, k\rangle=k|k, k\rangle\quad\hbox{and}\quad K_-|k,
k\rangle=0.\eqno(2.2)$$

When considering $|k, k\rangle$ as the fiducial state for the construction
of $\su$ CS, one can easily see that the latter are parametrized by
points of the coset space $\su/U(1)$.  More precisely, here one uses the
unit disc realization of $\su/U(1)$, namely $\d\cong\su/U(1)$,
which appears in the following construction of the CS [10],
$$|z\rangle\equiv
D^k_+(g)\equiv(1-\mz)^ke^{zK_+}|k, k\rangle.\eqno(2.3)$$   Note that
$\d=\{z\in\C\,\vert\enskip\mz<1\}$ and that $g$ in (2.3) belongs to $\su$
and projects down to $z\in\d$ through the principal bundle projection
$\su\longrightarrow\su/U(1)\cong\d$.  Moreover, $(1-\mz)^k$ in equation
(2.3) ensures that $\kz$ is normalized to one.  For short, the highest weight
state $|k, k\rangle$ will be, in the sequel, denoted $|0\rangle$.

According to Onofri [11], the unit disc can be equipped with an
$\su$-invariant symplectic (K\"ahler) $2$-form that will make it into a
classical phase space.  This is realized in the following way.  Define
first the K\"ahler potential as the phase space function
$$\eqalignno{f_0(z,\zb)&\equiv\log|\bo\kz|^{-2}\equiv
\log|\bo|D^k_+(g)\ko|^{-2}\cr
&=-2k\log(1-\mz).&(2.4)\cr}$$
Then a two-form on $\d$ is obtained as follows,
$$\omega_0\equiv-i\partial\bar\partial
f_0(z,\zb)=-2ik{dz\wedge d\zb\over(1-\mz)^2}\cdotp\eqno(2.5)$$
Clearly, $\omega_0$ is a closed and nondegenerate K\"ahler $2$-form.  In
(2.5) $\partial\equiv dz{\partial\over\partial z}$ and
$\bar\partial\equiv d\zb{\partial\over\partial \zb}$
such that the exterior derivative $d=\partial+\bar\partial$.

Having in hand the symplectic form we can now analyze the $\su$
action on $\d$.  The classical observables corresponding to the generators
of $\su$ given in (2.1) are nothing but the following Berezin covariant
symbols [10] [12],
$$K_0(z,\zb)\equiv\bz|K_0\kz=k{(1+\mz)\over(1-\mz)},\eqno(2.6a)$$
$$K_+(z,\zb)\equiv\bz|K_+\kz=2k{\zb\over(1-\mz)}\quad\hbox{and}\quad
K_-(z,\zb)\equiv\bz|K_-\kz=2k{z\over(1-\mz)}\cdotp\eqno(2.6b)$$
These functions generate a symplectic action of $\su$ on $\d$ through
Hamiltonian vector fields which are obtained using
the equation:
$$X_H\rfloor\omega_0=dH.\eqno(2.7)$$
Here the phase space function $H$ uniquely defines a vector field $X_H$
through the procedure of contraction (or interior product) with the
symplectic form $\omega_0$.  A straightforard calculation gives:
$$X_{K_0}=iz\pz-i\zb\pzb, \quad
X_{K_+}=i\pz-i\zb^2\pzb\quad\hbox{and}\quad
X_{K_-}=iz^2\pz-i\pzb\cdotp\eqno(2.8)$$
Because of (2.7) one clearly sees
that ${\cal L}_{X_H} \omega_0=0$ for $H=K_0, K_+$ or $K_-$, where ${\cal
L}_X$ is the Lie derivative along $X$.  This shows that the action of $\su$
on $\d$ is actually symplectic.  Moreover, $\omega_0$ defines in
$C^\infty(\d)$ a
Poisson bracket with respect to which the observables in (2.6) form a
symplectic realization of $\ssu$ in (2.1).  Indeed, we have
$$\{K_0, K_\pm\}\equiv i X_{K_0}\rfloor dK_\pm=\pm
K_\pm\quad\hbox{and}\quad\{K_+,K_-\} \equiv i X_{K_+}\rfloor
dK_-=-2K_0.\eqno(2.9)$$
Note that the factor $i$ appearing in the definition
of the Poisson bracket is inherent to the present complex-analytic context.

{}From equation (2.6) one shows that $\su/U(1)$ is a coadjoint
orbit of $\su$.  In other words, using (2.6) one defines the equivariant
momentum map,
$$\eqalignno{K:\ &\,\d\longrightarrow su^*(1,1)\equiv\ssu\cr
&\phantom{a}z\phantom{ab}\longmapsto(K_0(z,\zb), K_1(z,\zb),
K_2(z,\zb)),&(2.10)\cr}$$
where $K_+=K_1+iK_2$ and $K_-=K_1-iK_2$.  The three-vector on the
right hand side of
(2.10) spans a two dimensional surface in $\R^3$, the equation of which is
given by the
identity,
$$K_0^2(z,\zb)-K_1^2(z,\zb)-K_2^2(z,\zb)=k^2.\eqno(2.11)$$
For $k>0$ this is the upper component of the two-sheeted hyperbolo\"\i d in
$\R^3$ which
intersects the $K_0$ axis at the point $(k, 0, 0)$.  The unit disc $\d$ is
just the streographic
projection of the latter.  Thus $(\d, \omega_0)$ is a coadjoint orbit of $\su$.

\bigskip
\noindent{\bf 3. OSp(1/2) CS and the supersymplectic superunit disc}
\medskip \noindent
Now let us show how the previous constructions extend to the
supersymmetric world.  We will consider for instance the $\Osp$ CS
introduced in [7].  The $\osp$ Lie superalgebra has five supergenerators, $K_0,
K_\pm$ and $F_\pm$; their commutation-anticommutation relations
are as follows, $$[K_0, K_\pm]_-=\pm
K_\pm\quad\hbox{and}\quad[K_+, K_-]_-=-2K_0,\eqno(3.1a)$$
$$[K_0, F_\pm]_-=\pm{1\over2} F_\pm, \quad[K_\pm,
F_\pm]_-=0\quad\hbox{and}\quad[K_\pm, F_\mp]_-=\mp F_\pm.\eqno(3.1b)$$
$$[F_\pm, F_\pm]_+=K_\pm\quad\hbox{and}\quad[F_+, F_-]_+=K_0.\eqno(3.1c)$$
The commutator and the anticommutator are respectively denoted $[\,,\,]_-$
and $[\,,\,]_+$.  The use of the usual anticommutator
$\{\,,\,\}$ is here avoided in order to prevent any confusion with the
Poisson brackets.

It is quite remarkable that only the discrete series
representations of $\su$ extend to irreducible representations of $\Osp$
[16].  This observation encourages the study of the $\Osp$ CS as a
natural extension of those of $\su$.  Each irreducible representation (IR) of
$\Osp$ contains two IR of its subgroup $\su$ (see
(3.1a)).  More precisely, each IR of $\Osp$ labelled by $\tau$, such that the
Casimir operator $Q_2\equiv\tau(\tau-\demi)$, contains two discrete
series representations of $\su$, namely $D_+^{k=\tau}$ and
$D_+^{k=\tau+\demi}$.  The representation space is spanned by the set of
vectors $\left\{\, |\tau, k=\tau, m\rangle, |\tau, k=\tau+\demi, m\rangle,
m=k, k+1, \ldots\right\}$.  For more details concerning these
representations we refer to [7] [16] and references therein.

The fiducial state is chosen as before to be the ``highest weight'' state
$|\tau, k=\tau, m=\tau\rangle$ defined through the equations:
$$\eqalignno{K_0\,|\tau, k=\tau,
m=\tau\rangle&=\tau|\tau, k=\tau, m=\tau\rangle,&(3.2a)\cr
K_-|\tau, k=\tau, m=\tau\rangle&=0=F_-|\tau, k=\tau,
m=\tau\rangle.&(3.2b)\cr}$$
We will denote it $\ko$ as in section 2. The
$\Osp$ CS are then defined as  follows [7],
$$\kzt\equiv(1-\mz)^\tau\left[1-{1\over2}{\tb\t\over(1-\mz)}\right]^\tau
e^{zK_++\t F_+}\ko,\eqno(3.3)$$
here $z$ is a complex number which clearly belongs to $\d$ and $\t$ is an
odd coordinate, i.e. an anticommuting Grassmann number [1] [5].  Note that
$\zb$ is the usual complex conjugate of $z$ while $\tb$ is the so-called
adjoint
of $\t$ (denoted $\t^{\#}$ in [5]). This statement will be made clearer in
section
5.  Moreover, the expression in front of the exponential in (3.3) arises
from a normalization procedure as in section 2.  Because of (3.2) the pair
$(z,\t)$
parametrizes a realization of the $(2|2)$-dimensional homogeneous superspace
$\Osp/U(1)$. This defines the superunit disc $\dd\cong\Osp/U(1)$,
the global geometric structure of which will be discussed in next section.

We are now ready to extend Onofri's analysis to reveal the supersymplectic
character of $\dd$.  As in section 2 we use the SCS in (3.3) to first define
the following superpotential on $\dd$,
$$\eqalignno{f(z, \zb, \t,
\tb)&\equiv\log |\bo\kzt|^{-2}\cr
&=-2\tau\log(1-\mz)+\tau{\tb\t\over(1-\mz)},&(3.4)\cr}$$ from which we
extract the even two-superform $\omega$ on $\dd$ given by,
$$\eqalignno{\omega &\equiv-i\delta\bar\delta f(z, \zb, \t, \tb)\cr
&=-2i\tau\left[1+\demi\tb\t{(1+\mz)\over(1-\mz)}\right]{dz\wedge
d\zb\over(1-\mz)^2} +i\tau{d\t d\tb\over(1-\mz)}\cr
&\phantom{-2i\tau\left[1+\demi\tb\t{(1+\mz)\over(1-\mz)}\right]
{dz\wedge
d\zb\over(1-\mz)^2}}+i\tau\,\t\zb{dz\,d\tb\over(1-\mz)^2}
-i\tau\,z\tb{d\t\,d\zb \over(1-\mz)^2}\cdotp &(3.5)\cr}$$  In equation (3.5)
$\delta=dz\pz+d\t\pt$ and $\bar\delta=d\zb\pzb+d\tb\ptb$ such that the
exterior superderivative $d\equiv\delta+\bar\delta$.  Note also that we use
here Kostant's conventions [4], according to which the exterior superalgebra
on $\dd$ has a $\Z_+\times\Z_2$ bi-graded structure.  The $\Z_+$ gradation
is the usual gradation of exterior algebras while the $\Z_2$ one is the
natural gradation of the Grassmann algebra.  The
commutation-anticommutation relations of superforms are governed by the
following relation [4],
$$\beta_1\beta_2=(-1)^{a_1a_2+b_1b_2}\beta_2\beta_1,\eqno(3.6)$$ where
$a_1$ ($b_1$) is the degree of the superform $\beta_1$ with respect to the
$\Z_+$ ($\Z_2$) gradation.  For example, $dz\, d\zb=-d\zb\,dz$ (this is
clearly the usual wedge product), $dz\,d\tb=-d\tb\,dz$ and
$d\t\,d\tb=d\tb\,d\t$.  Using these conventions a straightforward
calculation shows that $\omega$ is closed, i.e. $d\omega=0$.  Hence, $\dd$ is
a supersymplectic supermanifold [1] [4] [14].  It is worth noting that
different conventions based on another grading of the exterior superalgebra
have been used by Berezin [1].  We will comment on this point in the
concluding section.

The Berezin covariant symbols or the classical observables corresponding to
the supergenerators of $\Osp$ have already been evaluated in [7].  They are
given by,
$$\openup 1mm\eqalignno{K_0(z, \zb,
\t,\tb)&\equiv\bzt|K_0\kzt=\tau{(1+\mz)\over(1-\mz)}\left[1+\demi{\tb\t\over(1-\
mz)}\right],
&(3.7a)\cr
K_+(z,\zb,\t,\tb)\!&\equiv\bzt|K_+\kzt=2\tau{\zb\over(1-\mz)}\left[1+\demi{\tb\t
\over(1-\mz)}\right],&(3.7b)\cr
K_-(z,\zb,\t,\tb)\!&\equiv\bzt|K_-\kzt=2\tau{z\over(1-\mz)}\left[1+\demi{\tb\t\o
ver(1-\mz)}\right],&(3.7c)\cr
F_+(z,\zb,\t,\tb)&\equiv\bzt|F_+\kzt=\tau{\tb+\zb\t\over(1-\mz)}\,,&(3.7d)\cr
F_-(z,\zb,\t,\tb)&\equiv\bzt|F_-\kzt=\tau{\t+z\tb\over(1-\mz)}\cdotp
&(3.7e)\cr}$$
Using equation (2.7), which applies equally well in the
supersymplectic context [4], one can evaluate the Hamiltonian vector
superfields corresponding to the observables given above.  A lenghty but
straightforward calculation gives:  $$\openup 2mm\eqalignno{
X_{K_0}&=iz\pz-i\zb\pzb+{i\over2}\t\pt-{i\over2}\tb\ptb\,,&(3.8a)\cr
X_{K_+}\!&=i\pz-i\zb^2\pzb-i\zb\tb\ptb\,,&(3.8b)\cr
X_{K_-}\!&=iz^2\pz-i\pzb+iz\t\pt\,,&(3.8c)\cr
X_{F_+}&={i\over2}\t\pz-{i\over2}\zb\tb\pzb-i\pt-i\zb\ptb\,,&(3.8d)\cr
X_{F_-}&={i\over2}z\t\pz-{i\over2}\tb\pzb-iz\pt-i\ptb\cdotp&(3.8e)\cr}$$
The same arguments which follow equation (2.6) still hold true here.  For
instance,  $\Osp$ acts, through the vector superfields found above, in a
supersymplectic way on $\dd$ and $\omega$ defines a Poisson superbracket
with respect to which the super observables in (3.7) form a supersymplectic
realisation of $\osp$.  For example, $$\eqalignno{\{K_0, F_\pm\}&\equiv
iX_{K_0}\rfloor dF_\pm=\pm{1\over2}F_\pm=-\{F_\pm,
K_0\},&(3.9a)\cr\{F_+, F_-\}&\equiv iX_{F_+}\rfloor dF_-=K_0=\{F_-,
F_+\}.&(3.9b)\cr}$$  As in the non-super case the superJacobi identities are
a direct consequence of the fact that $\omega$ is closed [1] [4].

As in section 2, equations (3.7) define a supermoment map $K^s$ as follows,
$$\eqalignno{K^s:\ &\,\dd\longrightarrow osp^*(1/2)\equiv\osp\cr
&\ (z,\t)\longmapsto(K_0(z,\zb,\t,\tb), K_i(z,\zb,\t,\tb),
F_i(z,\zb,\t,\tb)),\enskip
i=1,2,&(3.10)\cr}$$
where $K_+=K_1+iK_2$, $K_-=K_1-iK_2$, $F_+=F_1+iF_2$ and
$F_-=-i(F_1-iF_2)$.  (Notice here the use of the notion of adjoint of a
complex odd quantity, see section 5 for more details.)  The $(3|2)$-vector on
the right hand side of (3.10) spans a $(2|2)$-dimensional subsupermanifold
in $\R^{(3|2)}$, the equation of which is given by the identity,
$$K_0^2(z,\zb,\t,\tb)-K_1^2(z,\zb,\t,\tb)-K_2^2(z,\zb,\t,\tb)+
2F_1(z,\zb,\t,\tb)F_2(z,\zb,\t,\tb)=\tau^2.\eqno(3.11)$$ This is the
equation of an $\Osp$ supercoadjoint orbit.  It is the supersymmetric
extension of the upper component of the two-sheeted hyperbolo\"\i d of
section 2 (see eq. (2.11)). The superunit disc $\dd$ can then be viewed as a
superstereographic projection of the supercoadjoint orbit given in (3.11).
Hence, $\Osp/U(1)\cong(\dd,\omega)$ is a supercoadjoint orbit of $\Osp$.

Finally, let us mention that all the equations derived in this section reduce
to those of section 2 when one sets the odd parts to zero.  This completely
justifies our description of the results of this section as
supersymmetric extensions of those of section 2.  In particular, the body
of $\omega$ given in (3.5) is nothing but $\omega_0$
obtained in (2.5) with $k$ replaced by $\tau$.

\bigskip
\noindent{\bf 4.  Rothstein's theorem and the superK\"ahler character
of the superunit disc}
\medskip
\noindent We want now to go a step further and show that $(\dd,\omega)$
fits within the general picture of a supersymplectic supermanifold recently
depicted by Rothstein [14].  In order to make our statement understandable
a brief survey of known results is mandatory.  For more details see [1]-[4]
and [17].

A $(p|q)$-dimensional supermanifold is a pair $(M, \am)$ where $M$ is a
$p$-dimensional manifold with structure sheaf $\om$ and $\am$ is a sheaf
of supercommutative algebras ($\am$ is called the superstructure sheaf of
the supermanifold) such that: {\bf(a)} $\am/\nil$ is isomorphic to $\om$,
$\nil$ being the subsheaf of nilpotent elements of $\am$ and {\bf(b)}
$\she\equiv\nil/\nil^2$ is a locally free sheaf over $\om$ such that $\am$
is locally isomorphic to the exterior sheaf $\bigwedge\!\she$.  In other words
(b) means that given an open cover $\{U_\alpha\}$ of $M$, the following
map $\tau_\alpha$
$$\tau_\alpha: \am(U_\alpha)\longrightarrow
\om(U_\alpha)\otimes\hbox{$\bigwedge$}\R^q\equiv\hbox{$\bigwedge$}
\she(U_\alpha)\eqno(4.1)$$
is an isomorphism.  Here $\bigwedge\!\R^q$ is the exterior algebra on $\R^q$
and $q$ is the odd dimension of $(M,\am)$.  (In the complex case $\R$
should be replaced by $\C$.)  Local supercoordinates on $(M,\am)$ are
given by a set $(x^1,\ldots, x^p; \theta^1, \ldots, \theta^q)$ where
$(x^1,\ldots, x^p)$ are local coordinates on $M$ and
$(\theta^1, \ldots, \theta^q)$ form a basis of $\she$ over $\om$.

Batchelor's theorem [13] states that the local isomorphism in (4.1) extends
in a (non-canonical) way to a global one $\am\cong\bigwedge\!\she$.  In other
words, it shows that there exists a vector bundle $\F$ over $M$ such that
$\am$ is (non-canonically) isomorphic to $\Gamma(M, \bigwedge\!\F)$, the
sheaf of smooth sections of the exterior vector bundle
$\bigwedge\!\F\rightarrow M$.  More precisely this theorem applies only
when $M$ is a $C^\infty$ or a real-analytic manifold.  In these two cases the
supermanifold is said to be split.  When $M$ is a complex-analytic manifold
the theorem may not hold true.  This drawback shall not trouble us that
much since it is sufficient for our purpose to use the fact that one
can always associate a split supermanifold in the form $(M,
\Gamma(M,\bigwedge\!\F))$ to any supermanifold $(M,\am)$, for $M$ a
$C^\infty$, a real-analytic, or a complex-analytic manifold.  In fact in
practice one always deals with supermanifolds of that form.

Rothstein's theorem can be viewed as a supersymplectic extension of
Batchelor's theorem.  A supersymplectic supermanifold is defined as a
triple $(M,\am,\omega)$, where $\omega$ is a closed and non-degenerate
even $2$-superform on $(M,\am)$.  Rothstein's theorem allows one to
completely identify $\omega$ in terms of a symplectic structure on $M$
and extra structures in the vector bundle sector.  More precisely, it states
that to any supersymplectic supermanifold
$(M,\am,\omega)$ there corresponds a set $(M,\omega_0, \E, g, \nabla)$, where
$(M,\omega_0)$ is a symplectic manifold, $\E$ is a vector bundle over $M$
with metric $g$ and $g$-compatible connection $\nabla$, such that $\she$
is the sheaf of linear functionals on $\E$ and $\omega$ is completely
determined in terms of $(\omega_0,g,\nabla)$ as follows:
$$\omega=\omega_0+ \demi g_{ab}R^b_{ijc}\theta^a\theta^c dx^i dx^j
+g_{ab}D\theta^a D\theta^b.\eqno(4.2)$$
Here $R^b_{ijc}$, for $i,j\in\{1,\ldots,p=2n\}$ and $a,b\in\{1,\ldots,q\}$, are
the components of the curvature of $\nabla$, and $D$ is an operator
defined on $\bigwedge\!\she$ in terms of the components $A^a_{ib}$ of
$\nabla$, namely $$D\theta^a\equiv d\theta^a+A^a_{ib}\theta^b
dx^i\quad\hbox{and}\quad Dx^i=0.\eqno(4.3)$$  In the $C^\infty$ case this
correspondence is one-to-one [14].  Subsequently, {\it
Rothstein's data\/} will refer to the set $(M,\omega_0, \E, g, \nabla)$.

In what follows we will identify Rothstein's data for the supersymplectic
superunit disc $(\dd,\omega)$ obtained in the previous section.  This will
allow us to characterize $(\dd,\omega)$ using usual geometric structures.
Clearly $M$ in our case is nothing but the  unit disc $\d$ and $\omega_0$ is
its $SU(1,1)$-invariant symplectic form given in (2.5) with $k$
replaced by $\tau$. (From now on $\omega_0\equiv\omega_0(k=\tau)$.)  The
complete identification of Rothstein's data for $(\dd,\omega)$ is given in
the following theorem.
\medskip
\noindent{\bf Theorem:} {\sl The
$\Osp$-invariant supersymplectic structure of the superunit disc $\dd$  is
completely characterized by the set $\left(\d, \omega_0, T\d, \displaystyle
g\equiv\tau{dz\otimes d\zb\over(1-\mz)}, \nabla_g\right)$, where $T\d$
is the holomorphic tangent bundle of $\d$ and $\nabla_g$ is the (unique)
Hermitian connection associated to the (clearly) Hermitian and
$SU(1,1)$-invariant metric $g$.}\par
\medskip
\noindent{\it Proof}: The
proof is omitted since it consists of a straightforward verification. Using
the ingredients in the theorem one has to show that $\omega$ in (3.5) is
given by the formula (4.2).  In doing so one must make a small change in
(4.2), namely multiplying the terms at the right of $\omega_0$ by $i$.  This
is a natural change for the complex context. In fact, the formula
(4.2) was originally derived in the real case.  One can also easily verify that
$g$ is $SU(1,1)$-invariant, in other words that the vector fields $X_{K_0}$
and $X_{K_\pm}$ given in (2.8) are Killing vector fields for $g$, i.e. ${\cal
L}_{X_{K_0}}(g)=0={\cal L}_{X_{K_\pm}}(g)$.\qquad\qquad\qquad\boxit{5pt}{}

\medskip The metric $g$ above is in fact more than
Hermitian.  It is K\"ahlerian.  Indeed the associated $2$-form $\displaystyle
\omega_g\equiv -i\tau{dz\wedge d\zb\over(1-\mz)}$ on $\d$ is a K\"ahler
form since  $\omega_g$ is clearly closed.  Hence, the $\Osp$-invariant
supersymplectic structure $\omega$ on $\dd$ is completely determined
given the two K\"ahler structures $\omega_0$ and $\omega_g$ on the unit
disc $\d$.  This observation allows us to define the following general notion
of a superK\"ahler supermanifold.
\medskip
\noindent{\bf Definition:} {\sl A superK\"ahler
supermanifold $(M, \am, \omega)$ is a supersymplectic supermanifold, the
supersymplectic structure of which is completely determined given two
K\"ahler  $2$-forms $\omega_1$ and $\omega_2$ on $M$, such that
Rothstein's data is $(M, \omega_1, TM, g_{\omega_2},
 \nabla_{g_{\omega_2}})$, where $g_{\omega_2}$ is the K\"ahler metric on
$M$ associated to $\omega_2$ and $\nabla_{g_{\omega_2}}$ is the Hermitian
connection associated to $g_{\omega_2}$.}\par
\medskip
Alternatively one can define the superK\"ahler structure in terms
of K\"ahler potentials $f_1$ and $f_2$, which give rise to
$\omega_1$ and $\omega_2$, respectively.  For $(\dd,\omega)$,
$\omega_g$ can be obtained, as $\omega_0$ in (2.4)-(2.5), from a K\"ahler
potential $f_g$.  More precisely,
$$\omega_g=-i\partial\bar\partial f_g,\quad\hbox{where}\quad
f_g\equiv\tau Li_2(\mz) \eqno(4.4a)$$ and $Li_n(x)$ is the polylogarithmic
function defined by the series
$$Li_n(x)\equiv\sum_{k=1}^\infty {x^k\over
k^n},\quad 0\leq x\leq1. \eqno(4.4b)$$
$Li_2(x)$ is the so-called
dilogarithmic function.  Moreover, notice that the K\"ahler
potential $f_0$, which gives rise to $\omega_0$ in (2.5), is nothing but
$2\tau Li_1(\mz)$.  Hence, one can completely characterize $(\dd,\omega)$
by the triplet $(\d, f_0=2\tau Li_1(\mz), f_g=\tau Li_2(\mz))$.  It is worth
noting that a K\"ahler form does not specify uniquely a K\"ahler
potential.  Clearly, two K\"ahler potentials which differ only by a
holomorphic or/and antiholomorphic function produce the same K\"ahler
form.

In order to complete the present picture let us address the question of the
existence of a superK\"ahler potential.  We have a good candidate for this,
namely the superfunction $f(z,\zb,\t,\tb)$ on $\dd$ which is at the origin of
$\omega$ in (3.4)-(3.5).  This defines the notion of a superK\"ahler
potential.  It can actually be related to Rothstein's data.
Indeed, noting that $g_{z\zb}=\displaystyle
{\tau\over2}(1-\mz)^{-1}=g_{\zb z}$, one can rewrite (3.4) in the following
way:   $$f(z,\zb,\t,\tb)=f_0(z,\zb)+2g_{\zb
z}\,\tb\t.\eqno(4.5)$$
Or equivalently, using (4.4a),
$$f(z,\zb,\t,\tb)=f_0(z,\zb)+\partial_z\partial_{\zb} f_g(z,\zb)\,\tb\t.
\eqno(4.6)$$
 From this observation one can state the following Lemma.
\medskip
\noindent{\bf Lemma:} {\sl Rothstein's data for the superK\"ahler superunit
disc $\dd$ can be read off directly from the superK\"ahler potential $f$.}\par
\medskip
\noindent{\sl Proof:} It consists of direct computations based on
the observation made above. \boxit{5pt}{}
\medskip
\noindent On the other hand one can construct a superK\"ahler
potential for  $\omega$ in (3.5) given Rothstein's data as in the definition
above. These results generalize in a straightforward way to any superK\"ahler
supermanifold.

Let us close this section by a final remark.  As
previously stated, Rothstein's theorem allows the description of
supersymplectic supermanifolds in usual geometric terms.  It seems that one
can realize the invariance of supersymplectic forms under supergroups in
terms of the invariance of the associated Rothstein's data under usual groups
within the even part of the supergroup in question.  Actually, notice in our
case, the $\Osp$-invariance of $\omega$ in (3.5) versus the
$SU(1,1)$-invariance of both $\omega_0$ and $g$.  This
interesting question as well as those we have gathered in section 6 will be
addressed in a forthcoming publication.

\bigskip
\noindent{\bf 5. Darboux-Kostant coordinates} \medskip
\noindent
Darboux-Kostant's theorem [4] is the supersymmetric generalization of
Darboux's theorem.  Given a $(2n|q)$-dimensional supersymplectic
supermanifold $(M,\am,\omega)$, it states that for any open neighbourhood
$U$ of some point $m$ in $M$ there exists a set
$(q^1,\ldots,q^n,p_1,\ldots,p_n;$ $\xi^1,\ldots,\xi^q)$ of local coordinates on
$\bigwedge\!\she(U)$ such that $\omega$ on $U$ can be written in the following
form,
$$\omega\big\vert_U\equiv\tilde\omega=\sum_{i=1}^{n} dp_i\wedge dq^i+
\sum_{a=1}^q{\epsilon_a\over2}(d\xi^a)^2,\enskip
\epsilon_a=\pm1.\eqno(5.1)$$ This reflects the fact that any
supersymplectic supermanifold is locally a supersymplectic vector
superspace.  The coordinates in (5.1) are called Darboux-Kostant
coordinates.  Our aim in this section is to show that $(\dd,\omega)$
satisfies Darboux-Kostant's theorem.  To this end, we use the following
contraction-inspired trick.  (The full contraction $\Osp\rightarrow H(2|2)$,
$H(2|2)$ being
the super Heisenberg group of the extended plane $\R^{(2|2)}$, as well as
other contractions, will be addressed elsewhere.)
Instead of investigating locally $\dd$, we
introduce a length scale $r$ with respect to which the coordinates $z$ and
$\t$ become length-like quantities, then we take the limit
$r\rightarrow\infty$.  In other words, this procedure amounts to rescaling
$z$ in order to make the unit disc $\d$ into a disc of radius $\sqrt{2}r$.  In
the $r\rightarrow\infty$ limit, this unit disc becomes the whole plane.  For
consistency, one needs to rescale $\t$ also, such that in the same limit
$\dd$ becomes $\R^{(2|2)}$.  Hence, the rescaled coordinnates become
in the limit  $r\rightarrow\infty$ the Darboux-Kostant coordinates on
$\dd$.  Explicitly, let $z'$ and $\t'$ be the rescaled coordinates such that,
$$z'=\sqrt{2}\,r z\quad\hbox{and}\quad\t'=r\t.\eqno(5.2)$$ Take also
$\tau=r^2$.  Rewritting $\omega$ in
terms of $z'$ and $\t'$ and taking the limit $r\rightarrow\infty$ we find,
$$\tilde\omega\equiv\lim_{r\rightarrow\infty}{\omega(z',\t',\tau=r^2)}=
-idz'\wedge d\zb'+id\t' d\tb'.\eqno(5.3)$$
Clearly this is a closed
nondegenerate even $2$-superform on $\R^{(2|2)}$.  One needs now to
rewrite (5.3) in terms of ``real'' coordinates.  Let us define $q$, $p$,
$\xi^1$
and $\xi^2$ as follows,
$$z'={1\over\sqrt{2}}(q+ip),\quad \zb'={1\over\sqrt{2}}(q-ip),\eqno(5.4a)$$
and
$$\t'={1\over\sqrt{2}}(\xi^1+i\xi^2),\quad
\tb'=-{i\over\sqrt{2}}(\xi^1-i\xi^2).\eqno(5.4b)$$
Then,
$$\tilde\omega=dp\wedge dq+\demi(d\xi^1)^2+\demi(d\xi^2)^2,\eqno(5.5)$$
and is of the form (5.1). Finally, notice that in (5.4b) we have used the
(unusual) notion of the {\sl adjoint\/} of a complex Grassmann number
discussed in [5].  There it is denoted by a sharp ($\t^\#$) instead of a bar
($\tb$).  Hence, an odd Grassmann number is real
when $\t^\#\equiv\tb=-i\t$.  This is to be compared with the notion of
Hermitian conjugate of an odd quantum operator obtained by Rothstein as a
by product of  his quantization of a simple example of a supersymplectic
supermanifold [14].  The notion of reality we have used here is clearly
different
from that in [4].  On the other hand we could have considered the
usual notion
of reality and of complex conjugate, namely
$\overline{\xi^1+i\xi^2}\equiv\xi^1-i\xi^2$. In
doing so, one needs to use another form for (5.1).  More precisely,
$\epsilon_a=\pm i$ (see for instance [18]).

\bigskip
\noindent{\bf 6. Discussion, conclusions and prospects} \medskip
\noindent Here we discuss some of the results obtained in this work and
describe their possible generalizations.  We also briefly state other results.

\medskip
\noindent{\bf .OSp(N/2).}  First let us mention that $\dd\cong\Osp/U(1)$ is
not the only possible supersymmetric extension of the unit disc
$\d\cong\su/U(1)$.  In fact in the same way $\su$ admits several
supersymmetric extensions in the form of the orthosymplectic supergroups
$OSp(N/2)$, the unit disc possesses several susy extensions which are
homogeneous superspaces of the previous supergroups (we use here the fact
that $SU(1,1)$ is isomorphic to $Sp(2,\R)$).  More precisely, simple
arguments show that $OSp(N/2)/O(N)\times U(1)\cong{\cal D}^{(1|N)}$ is a
$(2|2N)$-dimensional supermanifold  that supersymmetrically extends $\d$.
The knowledge of $OSp(N/2)$ CS will make the explicit identification
of ${\cal D}^{(1|N)}$ easy.  As a conseqence notice that we should have called
$\dd$ the $N=1$ superunit disc.  The $N=2$ case will be considered in a
forthcoming publication.  This will provide us with a supersymplectic
supermanifold beyond the superK\"ahler setup.

\medskip
\noindent{\bf .OSp(N/2n).}  Clearly, superK\"ahler homogeneous
superspaces of $OSp(N/2)$ occur only when $N=1$.  This is the
superunit disc $\dd$.  A natural question then arises: when is a homogeneous
superspace of $OSp(N/2n)$ superK\"ahler? A purely dimensional argument
based on the known classification of K\"ahler homogeneous bounded domains
of type II (see [10] and references therein), which are the
homogeneous spaces $Sp(2n,\R)/U(n)\cong{\cal D}_{_{\!\hbox{\fiveti
II}}}^{({n^2+n\over 2})}$, helps one to conjecture the following:
$OSp(N/2n)/O(N)\times U(n)\cong{\cal D}_{_{\!\hbox{\fiveti
II}}}^{({n^2+n\over 2}|Nn)}$ is superK\"ahler when $n=2N-1$.  In other words,
${\cal D}_{_{\!\hbox{\fiveti II}}}^{(2N^2-N|2N^2-N)}\cong
OSp(N/4N-2)/O(N)\times U(2N-1)$ is a
$\left(2N(2N-1)|2N(2N-1)\right)$-dimensional superK\"ahler homogeneous
superspace of $OSp(N/4N-2)$ for $N\!\in\!\N$.  This superK\"ahler series
starts with the $N=1$ superunit disc $\dd$; for $N=2$ we obtain the
$(12|12)$-dimensional bounded superdomain ${\cal D}_{_{\!\!\hbox{\fiveti
II}}}^{(6|6)}$; etc.

\medskip
\noindent{\bf .Quantizations.} Because of the $\Osp$-invariance of its
supersymplectic structure, the superunit disc $\dd$ constitutes the perfect
arena for investigating or testing supersymmetric extensions of the known
quantization procedures.  The simplest one to consider in
the present context,  is clearly the CS-based one, namely the Berezin
quantization [12].  Extension of geometric quantization should also be tested
in the present situation.  With respect to this, another result of Rothstein
[14] is of great help in achieving the prequantization.  It states that the
expression at the right of $\omega_0$ in (4.2) is an exact superform, i.e.
$\omega=\omega_0+d\alpha$, $\alpha$ being an even $1$-superform.  This
allows one to prequantize $(M,\am,\omega)$ whenever
a prequantum map exists for $(M,\omega_0)$.  This is actually the case for
$(\d,\omega_0)$.  The prequantization of $(\dd,\omega)$ is then within
reach.  We will report on this elsewhere.  Note that the
(super)prequantization program is discussed in [4] and applied in [14] to a
simple example.  For completeness one should also consider $*$-quantization.

\medskip
\noindent{\bf .Berezin's conventions.} In the present work we have used
Kostant's conventions [4] (see eq.(3.6)). There
exist other conventions, namely Berezin's conventions.  A variant of these
latter is used in BRST formalism [18].  Berezin's conventions
are
based on the following two points: {\bf (a)} a simple grading of the
superalgebra
of exterior superforms which is induced from the Grassmann $\Z_2$ parity (the
Grassmann parity of an even (odd) $0$-form being $0$ $(1)$), and {\bf (b)} the
odd character of the exterior superderivative $d$.  This last point means that
when acting on a $p$-superform of Grassmann parity $\epsilon$, $d$ produces a
$p+1$-superform of parity $\epsilon+1$.  The Leibniz rule is as follows
$d(\omega_1\omega_2)=(d\omega_1)
\omega_2+(-1)^{\epsilon_{\omega_1}}\omega_1(d\omega_2)$.   Moreover, if
$r^A$ are coordinates of the supermanifold of parity $\epsilon_A$, we have [1]:
$r^Ar^B=(-1)^{\epsilon_A\epsilon_B} r^Br^A$,
$r^Adr^B=(-1)^{\epsilon_A(\epsilon_B+1)}dr^Br^A$ and $dr^Adr^B=
(-1)^{(\epsilon_A+1)(\epsilon_B+1)}dr^Bdr^A$.  For example
$dz\,d\zb=-d\zb\,dz$, $dz\,d\tb=d\tb\,dz$ and $d\t\,d\tb=d\tb\,d\t$.  Using
these conventions we can reproduce all the computations of section 3.  The
results are different, since Berezin and Kostant's exterior superalgebras are
not
isomorphic [3].  The closed two-superform turns out to be:
$$\eqalignno{\omega&=-2i\tau\left[1+\demi\tb\t{(1+\mz)\over(1-\mz)}\right]
{dz\wedge d\zb\over(1-\mz)^2} -i\tau{d\t d\tb\over(1-\mz)}\cr
&\phantom{-2i\tau\left[1+\demi\tb\t{(1+\mz)\over(1-\mz)}\right] {dz\wedge
d\zb\over(1-\mz)^2}}+i\tau\,\t\zb{dz\,d\tb\over(1-\mz)^2}
+i\tau\,z\tb{d\t\,d\zb \over(1-\mz)^2}\cdotp&(6.1)\cr}$$  However,
equation (2.7) is not consistent with Berezin's conventions.  In order to
be able to show that the observables given in (3.7) form a supersymplectic
realization of $\osp$, as was the case within Kostant's conventions in (3.9),
one
needs to modify equation (2.7) in the following way: $$X_H\rfloor
\omega\equiv\cases{ \phantom{-}dH&\hbox{when $H$ is even, i.e.
$\epsilon_H=0$}\cr  -dH&\hbox{when $H$ is odd, \ i.e.
$\epsilon_H=1$}\cr}\eqno(6.2)$$ Using this equation one shows that the
Hamiltonian vector superfields associated with the observables in (3.7) are
exactly those in (3.8).  Finally, equation (3.9) still holds true.

\medskip
\noindent{\bf .SuperMeasure.} One of the most important properties
of coherent states is the so-called resolution of identity [10].  It reflects
the fact that CS constitute an overcomplete basis of the Hilbert space.  The
explicit form of this property involves an integration over phase
space.  A measure on that space is then needed.  After lenghty
computations based on a generalization of Bogoliubov transformations, which
can  only be used when one considers the harmonic oscillator representation of
$\Osp$,
Balantekin {\sl et al.} were able to produce such a measure on $\dd$ [7].
Using
Berezin's notion of a density [1][3], our geometric approach allows us to
determine  this measure in a straightforward way.  Indeed, up to a
multiplicative constant, this measure is given by the formula,
$$d\mu(z,\zb,\t,\tb)\propto \sqrt{\hbox{\sl sdet}\,\Vert\omega_{AB}\Vert}\
{dz\,d\zb\over 2i\pi}\,d\t\,d\tb.\eqno(6.3)$$ Here ${\hbox{\sl
sdet}\,\Vert\omega_{AB}\Vert}$ is the superdeterminant or Berezenian of the
supermatrix $\Vert\omega_{AB}\Vert$, for $\omega$ given in (3.5) and
$\omega=dr^A\omega_{AB}dr^B$.  A direct computation gives,
$$d\mu(z,\zb,\t,\tb)\propto
{1\over\pi}{1\over(1-\mz)}\left[1+\demi{\tb\t\over(1-\mz)}\right]
dz\,d\zb\,d\t\,d\tb, \eqno(6.4)$$ which is the measure obtained in [7].
Starting with $\omega$ in (6.1) and repeating the above computations one
obtains the same measure on $\dd$, except from
the differences in the commutation-anticommutation relations of the
$1$-superforms.

\medskip
\noindent{\bf .Integral curves.} The integration of the flows of the
Hamiltonian vector superfields given in (3.8) amounts to solving a system of
nonlinear ordinary superdifferential equations. These are super-Riccati
equations, the resolution of which has been fully studied in [19].

\vfill\eject
\noindent{\bf Acknowledgements}
\medskip
\noindent The author gratefully thanks S.T. Ali for asking him the question
at the origin of the present work and for several stimulating discussions.
He also thanks V. Hussin for guiding his first steps into
supersymmetry, and L.M. Nieto for useful comments. Special thanks are due to
S.T.
Ali and V. Hussin for their kind hospitality at Concordia University and at the
Centre de Recherches Math\'ematiques of Universit\'e de Montr\'eal.

\bigskip\medskip
\noindent{\bf References}
\medskip
\leftskip 0.7cm

\lmp{1.}Berezin, F.A., {\it Soviet J. Nuclear Phys.} {\bf 29}, 857 (1979).

\lmp{2.}Leites, D.A., {\it Russian Math. Surveys\/} {\bf 13}, 1 (1980).

\lmp{3.}Manin, Y.I., {\it Gauge Field Theory and Complex Geometry\/},
Springer-Verlag, Berlin, 1988.

\lmp{4.}Kostant, B., in: {\it Lecture Notes in Mathematics vol.570\/}, 177,
(Bleuler, K. and Reetz, A. eds), Proc. Conf. on Diff. Geom. Meth.
in Math. Phys., Bonn 1975., Springer-Verlag, Berlin, 1977.

\lmp{5.}Cornwell, J.F., {\it Group Theory in Physics vol.III\/},
Academic Press, London, 1989.

\lmp{6.}Freund, P.G.O., {\it Introduction to Supersymmetry\/}, Cambridge
University Press, Cambridge, 1986.

\lmp{7.}Balantekin, A.B., Schmitt, H.A. and Barrett, B.R., {\it J. Math.
Phys.\/} {\bf 29}, 1634 (1988).

\lmp{8.}Balantekin, A.B., Schmitt, H.A. and Halse, P., {\it J. Math.
Phys.\/} {\bf 30}, 274 (1989).

\lmp{9.}Fatyga, B.W., Kostelecky, V.A., Nieto, M.M. and Truax, D.R., {\it Phys.
Rev.\/} {\bf D43}, 1403 (1991).

\lmp{10.}Perelomov, A., {\it Generalized Coherent States and Their
Applications\/}, Springer-Verlag, Berlin, 1986.

\lmp{11.}Onofri, E., {\it J. Math. Phys.\/} {\bf 16}, 1087 (1975).

\lmp{12.}Berezin, F.A., {\it Commun. Math. Phys.} {\bf 40}, 153 (1975).

\lmp{13.}Batchelor, M., {\it Trans. Amer. Math. Soc.} {\bf 253}, 329 (1979).

\lmp{14.}Rothstein, M., in: {\it Lecture Notes in Physics vol.375\/},
331 (Bartocci, C., Bruzzo, U., and Cianci, R., eds), Proc. Conf. on Diff.
Geom. Meth.
in Math. Phys., Rapallo 1990., Springer-Verlag, Berlin, 1991.

\lmp{15.}Klauder, J.R. and Skagerstam, B.-S., {\it Coherent
States - Applications in Physics and Mathematical Physics\/}, World
Scientific, Singapore, 1985.

\lmp{16.}Hughes, J.W.B., {\it J. Math. Phys.\/} {\bf 22}, 245 (1981).

\lmp{17.}Shnider, S. and Wells, R.O.Jr., {\it Supermanifolds, Super Twistor
Spaces and Super Yang-Mills Fields}, Les Presses de l'Universit\'e de
Montr\'eal, Montr\'eal, 1989.

\lmp{18.}Henneaux, M., {\it Classical Foundations of BRST Symmetry\/},
Bibliopolis, Napoli, 1988; and Henneaux, M. and Teitelboim, C., {\it Commun.
Math.
Phys.\/} {\bf 115}, 213 (1988).

\lmp{19.}Beckers, J., Gagnon, L., Hussin, V. and Winternitz, P., {\it J. Math.
Phys.\/} {\bf 13}, 2528 (1990).

\end